\documentclass[11pt,a4paaper,oneside]{article}
\usepackage{lineno,hyperref}

\usepackage{graphicx}              
\usepackage{epsfig}
\usepackage{amsmath}               
\usepackage{amssymb}
\usepackage{epstopdf}
\usepackage{verbatim}
\usepackage{mathptmx}
\usepackage{mathalpha}
\usepackage{mathtools}
\usepackage[scale=0.80]{geometry}
\usepackage{graphicx, times}
\usepackage{amsfonts}              
\usepackage{amsthm}                
\usepackage{multicol}
\usepackage{algorithm}
\usepackage{algorithmic}
\usepackage{varwidth}
\usepackage{parskip}
\usepackage{hyperref}
\usepackage{rotating}
\usepackage[numbers,sort&compress]{natbib}
\usepackage{multirow}
\usepackage{pdflscape}
\usepackage[numbers]{natbib}
\usepackage{caption}
\usepackage{xcolor}
\usepackage{color}
\usepackage{comment}
\usepackage{subcaption}
\newcommand*{\rom}[1]{\expandafter\@\romannumeral #1}

\newcommand{\bea}{\begin{eqnarray}}
	\newcommand{\eea}{\end{eqnarray}}
\newcommand{\bee}{\begin{eqnarray*}}
	\newcommand{\eee}{\end{eqnarray*}}

\begin{document}
\author{Romanshu Garg$^{1}$\footnote{romanshugarg18@gmail.com }, G. P. Singh$^{1}$\footnote{gpsingh@mth.vnit.ac.in}, 
\vspace{.3cm}\\
${}^{1}$ Department of Mathematics,\\ Visvesvaraya National Institute of Technology, \\ Nagpur 440010, Maharashtra, India.
}
\date{}

\title{The $f(Q, T)$ gravity and affine EoS: observational aspects  }

\maketitle

\begin{abstract}
In this paper, we investigate the cosmic expansion scenarios within the framework of $f(Q, T)$ gravity  by using the affine equation of state(EoS) parameter. Specifically, we consider the linear form of $f(Q, T)= Q+ \beta T $, where $\beta$ is a free model parameter. We use the Bayesian statistical methods, specifically the $\chi^{2}$ minimization technique to constrain the model parameters using the Cosmic Chronometer (CC) data, Pantheon+SH0ES and DESI BAO data. We further analyse the characteristics of the derived cosmological model. A comprehensive study of the energy density, pressure, equation of state parameter and cosmographic parameters are carried out to understand the universe's evolution in the model. The determination of the universe’s present age for this model is within $1\sigma$ with the plank results.
\end{abstract}
\section{Introduction}\label{sec:1}
Current astronomical data strongly indicate that the universe is expanding at an accelerating rate in the present epoch \cite{1998AJ....116.1009R,1999ApJ...517..565P, 2020A&A...641A...6P}. The component responsible for this stage of the universe’s expansion is commonly called dark energy (DE), which currently accounts for approximately $68\%$ of the universe’s energy density. A comprehensive discussion of the theoretical background of dark energy and its various models can be found in several excellent review articles \cite{sahni2000case, padmanabhan2003cosmological}. Till now, the true nature of dark energy (DE) remains unknown. In this context, various dark energy (DE) models have been proposed to align with recent observational data and the $\Lambda$CDM model is considered the simplest suitable models among these models in this series. According to this model, dark energy is explained as a constant vacuum energy density, interpreted as the cosmological constant. This model aligns with the observed data. Despite its success, this model faces challenges known as the fine tuning problem and the coincidence problem \cite{weinberg1989cosmological,
steinhardt1999cosmological, di20realm}. For an in-depth analysis of the issues surrounding the cosmological constant model, refer to the reviews by Padmanabhan \cite{padmanabhan2003cosmological} and Carroll \cite{carroll2001cosmological}. To resolve these issues, considering alternative possibilities to explain the origin and nature of DE is a logical approach. Consequently, it is necessary to thoroughly analyze various approaches and extensions of GR. A variety of techniques have been introduced in the past few decades to address modern cosmological issues. Nowadays, the most widely accepted solution to existing problems is based on the modified gravity theory approach. The $f(R)$ gravity \cite{buchdahl1970non}, a modification of General Relativity, is one of the most well-known approaches to addressing the dark content problem of the cosmos. Over the years, multiple modified theories \cite{harko2011f,
bamba2010finite,nojiri2011unified, nojiri2017modified,capozziello2023role,
jimenez2018coincident,capozziello2019extended,cai2016f,
singh2024lyra,singh2021unified,singh2024abcde,garg2025late,
kotambkar2017anisotropic,singh2025dynamical,
chakraborty2024existence,chakraborty2025exploring} have been subjected to research and analysis.
\par Recently, Xu et al. \cite{xu2019f} introduced a novel extension of $f(Q)$ gravity, incorporating a non-minimal coupling between the nonmetricity function $Q$ and $T$, where $T$ represents the trace of the matter energy momentum tensor. In a more precise sense, the Lagrangian density of the gravitational field is defined as a general function of $Q$ and $T$. In their proposal, Xu et al.  \cite{xu2019f} predicted a de Sitter-type expansion of the universe and introduced a geometric alternative to dark energy (DE). Indeed, this modified theory is currently gaining attention with several intriguing studies emerging in the literature. In  \cite{zia2021transit,pradhan2021models} the $f(Q,T)$ gravity theory with quadratic $Q$ and linear $T$, as well as its generalized form has been analyzed to derive constraints on the model parameters using the $R^2$ test formula. Additionally, various aspects, including Cosmological inflation \cite{SHIRAVAND2022101106} and Cosmological perturbations \cite{najera2022cosmological} have been extensively studied. Numerous further investigations have been carried out within the framework of $f(Q, T)$ gravity theory \cite{lalke2023late,singh2022cosmological, mandal2023cosmic, xu2020weyl, najera2022effects,shukla2025late, shekh2023new,maurya2026constrained,mohanty2026reconstruction}. Motivated by the above work, we aim to examine the cosmological viability of the modified $f(Q, T)$ gravity theory in explaining the current and late-time acceleration of the universe. 
\par In this manuscript, we investigate the cosmological implications of an affine equation of state (EoS) within the framework of $f(Q, T)$ gravity. The affine EoS characterizes both hydrodynamically stable and unstable fluid regimes while maintaining a constant sound speed. The affine equation of state(EoS) can be described by the relation  $p= n \rho -m$, where $m$ and $n$ are constant parameters. According to the classical stability condition \cite{peebles2003cosmological,ellis2007causality} the parameter $n$ must satisfy $0 \leq n \leq 1$. This affine EoS exhibits a well defined the $\Lambda$CDM limit in the low-energy regime \cite{ananda2006cosmological}. Additionally, In non-conservative gravity models, this form of equation of state may arise inherently near the bouncing epoch of the universe \cite{singh2022cosmic}. The inflationary dynamics associated with this equation of state can lead to a constant-roll inflationary phase and  the late-time evolution of the universe exhibits accelerated expansion consistent with the generalized second law of thermodynamics \cite{singh2018thermodynamical}. This affine equation of state has been extensively investigated in both cosmological and astrophysical contexts across various gravitational frameworks \cite{chiba1997cosmology, babichev2004black, babichev2005dark, vstefanvcic2005expansion, balbi2007lambda, singh2008bianchi, chaubey2009bianchi, khadekar2015brane}. Motivated by these distinctive features, we investigate the cosmological consequences of the affine equation of state within the framework of $f(Q, T)$ gravity.

\par The arrangement of this manuscript is as follows: A fundamental introduction of $f(Q, T)$ gravity and the cosmological solution for the affine EoS model within the framework of $f(Q, T)$ gravity are presented in section (\ref{sec:2}). In Section (\ref{sec:5}), the cosmological solution's compatibility is examined with Cosmic Chronometer (CC) and Joint (CC+Pantheon+SH0ES+DESI BAO) datasets using Bayesian statistical techniques. A detailed analysis of cosmological dynamics including energy density, pressure, cosmographic parameter and age of the universe are discussed in section (\ref{sec:6}). In section (\ref{sec:7}), we summarize our results.
\section{Mathematical formalism in $ f(Q, T)$ gravity}
\label{sec:2}
In the framework of extended symmetric teleparallel gravity, the action is formulated with the matter Lagrangian $L_m$ is described in \cite{xu2019f}.
\begin{equation}{\label{1}}
S = \int \sqrt{-g}  \, d^4x \left[\frac{1}{16 \pi}f(Q,T) + \mathcal{L}_m\right]. 
\end{equation}
In this context, $Q$ signifies the non-metricity scalar whereas $T$ represents the trace of the energy-momentum tensor and $g = det(g_{ab})$. According to Jimenez et al. \cite{jimenez2018coincident}, the non-metricity scalar is defined as follows
\begin{equation}{\label{2}}
Q \equiv -g^{ab} \left( L^{\alpha}_{\beta a}L^{\beta}_{b \alpha} -L^{\alpha}_{\beta \alpha}L^{\beta}_{ab}\right).
\end{equation}
In this context, $L^{\alpha}_{\beta \gamma}$ denotes the disformation tensor which is defined as:
\begin{equation}{\label{3}}
L^\alpha_{\beta \gamma} \equiv -\frac{1}{2} g^{\alpha c} \left( \nabla_{\gamma} g_{\beta c} + \nabla_{\beta} g_{c \gamma} - \nabla_{c} g_{\beta \gamma} \right).
\end{equation}
Within the framework of symmetric teleparallel gravity, the non-metricity tensor $Q_{\gamma ab}$ is given as the covariant derivative of the metric tensor $g_{ab}$ concerning the Weyl-Cartan connection i.e. \( Q_{\gamma ab} = \nabla_{\gamma} g_{ab} \) and its traces are  \( Q_{\alpha} \equiv Q_{\alpha \ a}^{\ a} \), \( \tilde{Q}_{\alpha} \equiv Q^{a}_{\ \alpha a} \).
\vspace{0.2cm}\\
According to \cite{xu2019f}, the superpotential of the model is 
\begin{equation}{\label{4}}
P^\alpha_{ab} = -\frac{1}{2} L^\alpha_{ab} + \frac{1}{4} (Q^{\alpha} - \tilde{Q}^{\alpha}) g_{ab} - \frac{1}{4} \delta^\alpha_{\ (a} Q_{b)},
\end{equation}
and utilizing the non-metricity scalar and the disformation tensor, we can derive the following relation.
\begin{equation}{\label{5}}
Q = -Q_{\alpha ab} P^{\alpha ab}.
\end{equation}
In addition, the energy-momentum tensor $T_{ab}$ and its variation concerning $g_{ab}$ can be described as follows.
\begin{equation}{\label{6}}
T_{ab} \equiv -\frac{2}{\sqrt{-g}} \frac{\delta \left( \sqrt{-g} \mathcal{L}_m \right)}{\delta g^{ab}}, \quad  \frac{\delta g^{\alpha\beta} T_{\alpha\beta}}{\delta g^{ab}}  = T_{ab} + \Theta_{ab},
\end{equation}
where
\begin{equation}{\label{7}}
\Theta_{ab} \equiv g^{\alpha\beta} \frac{\delta T_{\alpha\beta}}{\delta g^{ab}}.
\end{equation}
\begin{equation}{\label{8}}
8\pi T_{ab} = -\frac{2}{\sqrt{-g}} \nabla_{\alpha} \left( f_Q \sqrt{-g} P^{\alpha}_{\ ab} \right) 
-\frac{1}{2} f g_{ab} + f_T (T_{ab} + \Theta_{ab}) 
- f_Q \left( P_{a\alpha\beta} Q_{b}^{\ \alpha\beta} - 2Q_{a}^{\ \alpha\beta} P_{\alpha\beta b} \right).
\end{equation}
\vspace{0.1cm}\\
In the entire analysis, we consider a spatially flat FLRW universe \cite{partridge2004introduction} whose metric is given by 
\begin{equation}{\label{9}}  
ds^{2}=-dt^{2}+a^{2}(t) \left( dx^{2}+ dy^{2}+ dz^{2}\right),
\end{equation}
where $H\equiv \frac{\dot{a}}{a}$ denotes the Hubble parameter, $a(t)$ represents the scale factor. The non-metricity scalar $Q$ is computed as $Q=6H^{2}$, as well explained in ref. \cite{jimenez2018coincident}.\\
We assume that the universe's matter content consists of a perfect fluid whose energy-momentum tensor can be written as $T^{a}_{b}=diag(-\rho, p, p, p)$, where cosmic pressure denoted by  $p$ and $\rho$ represents the energy density. Using the FLRW metric (\ref{9}) along with Eqs. (\ref{4}), (\ref{6}), (\ref{7}) and (\ref{8}), we can express as follows.
\begin{equation}{\label{901}}
8\pi \rho + 8\pi \tilde{G}(\rho + p)=\frac{f}{2} - 6F H^2 
\end{equation}
\begin{equation}{\label{902}}
-8\pi p=\frac{f}{2} - 2\left[F\left(\dot{H} + 3H^2\right)+\dot{F}H  \right]  
\end{equation}
To simplify the expressions, we use the notations \( F \equiv f_Q \) and \( 8\pi \tilde{G} \equiv f_T \) and the terms $f_Q$ and $f_T$ represent the differentiation of the function $f$ with respect to $Q$ and $T$ respectively. From the above Eqs. (\ref{901}) and (\ref{902}), one can derive the following expressions.
\begin{equation}{\label{903}}
8\pi \rho =  - \frac{2\tilde{G}}{1 + \tilde{G}}\left(\dot{F}H + F\dot{H}\right) + \frac{f}{2} - 6FH^2
\end{equation}
\begin{equation}{\label{904}}
8\pi p =  2\left(\dot{F}H + F\dot{H}\right)-\frac{f}{2} + 6FH^2 
\end{equation}
To describe the transitional evolution of the universe in this model, we consider  $\rho$ and $p$ as the energy density and pressure corresponding to the affine equation of state \cite{babichev2005dark,balbi2007lambda,
singh2024affine}
\begin{equation}{\label{701}}
p=n\rho-m.
\end{equation}
Where, the parameters $m$ and $n$ are constant parameters and estimated by fitting the model with standard cosmological observations. Notably, $n$ is a dimensionless quantity while $m$ possesses the dimensions of energy density. For $m=0$, the equation of state reduces to the standard form $p=\omega \rho$, where $n=\omega$. The parameter $n$ can be interpreted as a constant sound speed and the classical stability criterion imposes the constraint $0 \leq n \leq 1$. Issues related to causality and stability play a crucial role in determining the physical viability of cosmological models \cite{peebles2003cosmological,ellis2007causality}. Equation (\ref{701}) provides a generalized framework capable of explaining quintessence as well as phantom dark energy with the adiabatic sound speed squared defined as $c^{2}_{s}\equiv \frac{\partial p}{\partial \rho}=n \geq 0$. The causality principle further imposes the constraint $0 \leq c_{s}^{2} \leq 1$. Furthermore, the affine  equation of state (\ref{701}) admits a well defined $\Lambda$CDM limit characterized by $p_{\Lambda}=-\rho_{\Lambda}$. Consequently, the proposed EoS can describe a dynamical dark energy scenario with a slowly varying behavior at late times, depending on the chosen model parameters. In this framework, the continuity equation is written as 
\begin{equation}{\label{690}}
\dot{\rho} +3H(\rho+p)=0.
\end{equation}
To examine the compatibility of the cosmological solution with observational data, it is convenient to express the relevant quantities in terms of redshift. Using the scale factor–redshift relation $\frac{a_{0}}{a(t)}=1+z$ with $a_{0}=1$, the time derivative can be transformed as $\frac{d}{dt}=-(1+z)H(z)\frac{d}{dz}$. Using  Eq. (\ref{903}), (\ref{904}), (\ref{701}) and (\ref{690}) the differential equation yields the solution.
\begin{equation}{\label{601}}
H(z)=H_{0}\sqrt{ \frac{(1+z)^{\left(\frac{6(n+1)(8\pi +\beta)}{16\pi +3\beta-\beta n}   \right)   }  }{(m+1)} +  \frac{  \frac{m}{(1+n)}-\frac{\beta m}{\beta n -3\beta -16\pi}   }{(m+1)} }    
\end{equation}
Here, $H_{0}$ represents the present day value of the Hubble parameter at $z=0$. The observational constraints on the parameters $H_{0}, \beta, m $ and $n$ provide a measure of the model’s consistency with observational data as well as its compliance with the causality condition.

\section{Observational Constraints}\label{sec:5}
This section focuses on describing the data sets that utilized to find out the median values of model parameters of the model. In this work, we conduct a Bayesian study to determine the observational viability of the current cosmological model. To determine the parameter constraints for $H_0$, $\beta$, $m$ and $n$, we analyze two observational datasets, namely the cosmic chronometer (CC) and the Joint (CC+Pantheon+SH0ES+DESI BAO) dataset. The $\chi^{2}$ minimization method along with the Markov Chain Monte Carlo (MCMC) technique is used for statistical analysis implemented with the emcee tool~\cite{foreman2013emcee}.

\subsection{The Cosmic chronometer data}
\label{sec:5.1}
\begin{center}
\begin{figure}
\includegraphics[width=15.5cm, height=7.5cm]{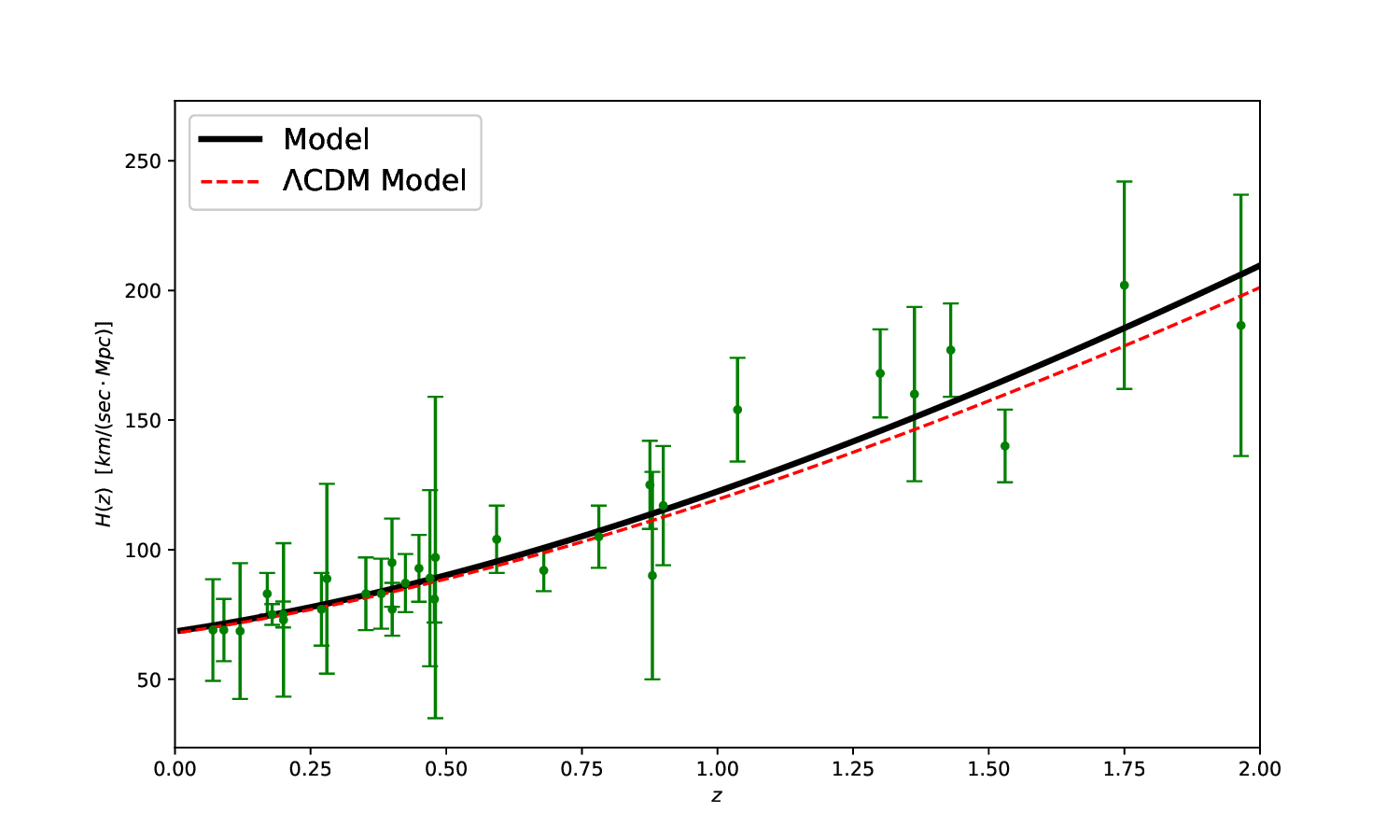}
\caption{In comparison to the $\Lambda CDM$ model, the best fit Hubble parameter versus $\mathit{z} $.}
\label{fig:1}
\end{figure}
\end{center}

The expansion rate of the universe is determined by the Hubble parameter at any instant and its observational values play a key role in studying dark energy and cosmic evolution. The model parameters ($H_0$, $\beta$, $m$ and $n$) are constrained using Cosmic Chronometer data set which include $31$ data points \cite{28,simon2005constraints} measured through the differential ages of galaxies method in the redshift range $0.07 \leq z \leq 1.965$. Jimenez and Loeb \cite{jimenez2002constraining} introduced the fundamental principle of cosmic chronometer observations, which relates the Hubble parameter $(H(z))$, cosmic time $(t)$ and redshift$(z)$ as $H(z)=\frac{-1}{(1+z)}\frac{dz}{dt}$. Observational constraints on the parameters $H_0$, $\beta$, m and $n$ are obtained by minimizing the $\chi^{2}$ function, (which corresponds to the maximization of the likelihood function), as discussed in \cite{lalke2024cosmic,mandal2024late, mandal2023cosmic,singh2024affine, singh2025observational}:
\begin{equation}{\label{24}}
\chi^{2}_{CC}(\theta)=\sum_{i=1}^{31} \frac{[H_{th}(\theta,z_{i})-H_{obs}(z_{i})]^{2}    }{ \sigma^{2}_{H(z_{i})}}.   
\end{equation} 
In this context, $H_{th}$ signify the theoretical values of the Hubble parameter, $H_{obs}(z_i)$ is used to signify the observed values. The notation $\sigma_{H}$ is assigned to express the standard deviation of each $H_{obs}(z_i)$ observed value.
\vspace{0.1cm}\\
The best-fit Hubble parameter curve derived from equations (\ref{601}) is displayed in figure $(\ref{fig:1})$ along with the error bars of the CC points. It can be observed that the cosmic dynamics of the model deviate from the $\Lambda$CDM model in the past, particularly for redshift $z > 1$. The close agreement between the model (black line) and $\Lambda$CDM (red dashed line) indicates that the proposed model is consistent with standard cosmology over the observed redshift range. The model provides a good fit to observations. It is observationally viable. At low redshift $(z \leq 1)$, propose cosmological model nearly overlap with $\Lambda$CDM Model which indicating similar predictions for the present day Universe. This figure demonstrates that the proposed cosmological model successfully reproduces the observed evolution of the Hubble parameter and remains in close agreement with the $\Lambda$CDM paradigm.
\subsection{Pantheon+SH0ES data}\label{sec:5.2}
Type Ia Supernovae (SNe Ia) play a crucial role in understanding the evolutionary history of the Universe. Observations of SNe Ia have been fundamental in confirming the accelerated expansion of the cosmos \cite{1998AJ....116.1009R,1999ApJ...517..565P}. In this work, we employ the Pantheon+SH0ES dataset \cite{PSH0ES}, which consists of $1550$ independent SNe Ia and a total of $1701$ light-curve measurements spanning the redshift interval $0.00122 < z_{HD} < 2.26137$ \cite{brout2022pantheon+}. Compared to the earlier Pantheon compilation \cite{scolnic2018complete}, this dataset provides a significantly larger sample. The Pantheon+SH0ES(PS) dataset is particularly useful for constraining cosmological parameters as well as the present-day Hubble constant $(H_0)$. By incorporating Cepheid-based distance moduli along with the observed apparent magnitudes of SNeIa, it enables an independent determination of the absolute magnitude $M$ defined as $ M=m_{b}-\mu_{Ceph}$. This approach effectively removes the degeneracy between $M$ and $H_{0}$, allowing both parameters to be simultaneously estimated \cite{10.1093/mnras/stad451}. In our analysis, we adopt the apparent magnitude ($m_{b}$) as given in \cite{singh2025observational}
\begin{equation}{\label{7a}}
m_{b}=M+5\log_{10}\left(\frac{d_{L}(z)}{1  Mpc}\right)+25,
\end{equation}
where the luminosity distance $(d_{l})$ is given by
\begin{equation}{\label{1001a}}
d_{L}(z)=c(1+z)\int_{0}^{z}\frac{dz'}{H(z')}.
\end{equation}
$d_{l}$ has the ‘Length’ dimension. In the likelihood
analysis, we defined \cite{singh2025observational}
\begin{equation}{\label{1001b}}
\chi^{2}_{\mathrm{ps}} = \Delta Q^{T} C^{-1} \Delta Q
\end{equation}
The vector $Q$ is constructed from $ i_{th} $ elements, which is defined in \cite{10.1093/mnras/stad451}
\begin{equation}{\label{1001c}}
Q_i =
\begin{cases}
m_{bi} - M - \mu_{i\mathrm{Cep}}, & \text{if } i \in \text{Cepheid hosts}, \\
m_{bi} - M - \mu_{\mathrm{model}}(z_i), & \text{otherwise},
\end{cases}
\end{equation}
where $m_{bi} - M \equiv \mu_i$ will represent the distance modulus of $i_{th}$ SneIa and $\mu_{i\mathrm{Cep}}$ will represent the distance modulus of Cepheid host of $i_{th}$ SneIa. The $\mu_{i\mathrm{Cep}}$ is calculated independently \cite{riess2022comprehensive}. The matrix $C^{-1}$ denotes an inverse of the covariance matrix and it consists of both the statistical and systematic uncertainties. The theoretical distance modulus is given by 
\begin{equation}{\label{1001d}}
\mu_{model}=5\log_{10}\left(\frac{d_{l}}{Mpc}\right)+25.
\end{equation}
Here $d_{l}$ is in the unit of Mpc. In this study, the parameter $M$ is estimated along with the other cosmological parameters. It is important to note that the value of $M$ is not fixed and can change depending on the cosmological model under consideration \cite{deliduman2024f,singh2025observational, 10.1093/mnras/stad451}. The associated chi-square is expressed as $\chi^{2}_{PS}$.
\subsection{DESI BAO data}\label{sec:5.3}
\begin{center}
\begin{figure}
\includegraphics[width=21.0cm, height=19.5cm]{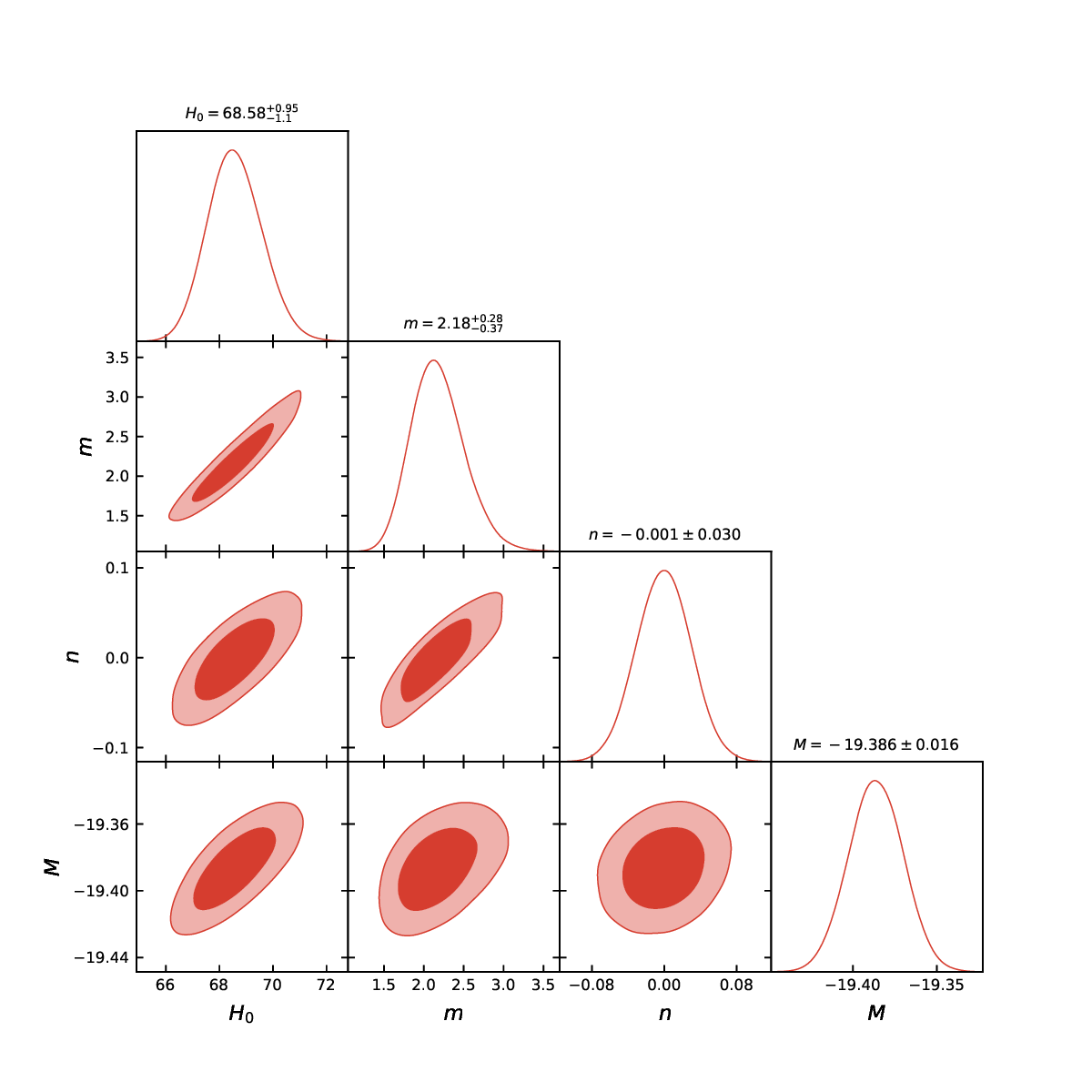}
\caption{Marginalized $1D$ and $2D$ contour map with median values of $H_{0}$, m and n using Joint data set.}
\label{fig:2}
\end{figure}
\end{center}

The baryonic acoustic oscillation (BAO) measurements provide crucial information related to the baryonic decoupling redshift $z_{d} \approx 1059$. The Dark Energy Spectroscopic Instrument (DESI) delivers BAO measurements over a wide redshift interval of $0.1 < z < 2.4$. These observations give information about the quantities $D_{H}/r_{d}$, $D_{M}/r_{d}$ and $D_{v}/r_{d}$, where the Hubble distance is defined as $ D_{H} \equiv c/H$, the transverse comoving distance as
$D_{M} \equiv c\int_{0}^{z} dz'/H(z')$ and the angle averaged distance as $ D_{v} \equiv (zD^{2}_{M} D_{H})^{\frac{1}{3}}$. Here, $r_d$ represents the comoving sound horizon at the redshift drag redshift $z_d$. In this work, we adopt the data vector, analysis framework and terminology presented by Adame et al. \cite{adame2025desi} and Li et al. \cite{li2025comprehensive}. The associated chi-square is expressed as $\chi^{2}_{BAO}$ and we take $r_{d}=147.09$ \cite{adame2025desi}.
\vspace{0.3cm}\\
We employ the emcee package \cite{foreman2013emcee} along with equation (\ref{601}) to obtain the maximum likelihood estimate using the Joint (CC+Pantheon+SH0ES+DESI BAO) dataset. The Joint $\chi^{2}$ employed in maximum likelihood analysis is defined as the combined $\chi^{2}_{CC}+\chi^{2}_{PS}+\chi^{2}_{BAO}$. Contour map at the $1\sigma$ and $2\sigma$ confidence level together with the $1D$ posterior distribution for Model is displayed in figure $(\ref{fig:2})$.  Table $(\ref{table:1})$ present the median values of the model parameters (obtained from \textit{emcee} using the MCMC analysis) for the Hubble parameter (\ref{601}). The parameter estimates summarized in Table $(\ref{table:1})$ indicate that the model is consistent with observational data.
\par For this model, the $1\sigma$ confidence interval for the parameter $n$ is found to be $(0.014, 0.098)$ based on CC data. When combined with the Pantheon dataset (with $1048$ data points) (CC+Pantheon dataset) the median value of $n$ is $n=0.108^{+0.074}_{-0.094}$ with the $1\sigma$ range
$(0.014, 0.182)$. These constraints indicate that $n$ remains positive throughout the parameter space ensuring classical stability within the interval $0 \leq n \leq 1$. Hence, the model is classically stable for both CC and CC+Pantheon datasets. In contrast, for the extended dataset including SH0ES and DESI BAO, i.e., CC+Pantheon+SH0ES+DESI BAO, the median value of $n$ is obtained  $-0.001\pm 0.03 $ with a $1\sigma$ range of $(-0.031, 0.029)$. In these cases, the results show that within the $1\sigma$ confidence level, the parameter $n$ attains negative values, implying that the model may not be classically stable for these values. Nevertheless, a significant portion of the parameter space remains positive, satisfying the classical stability condition $0 \leq n \leq 1$. This indicates that there exist certain values of the parameter $n$ for which the model remains classically stable for the CC+Pantheon+SH0ES+DESI BAO datasets. Therefore, the model cannot be regarded as completely unstable subjected to the CC+Pantheon+SH0ES+DESI BAO datasets. In summary, the model exhibits classical stability for CC and CC+Pantheon ($1048$ data points) while for the extended dataset (CC+Pantheon+SH0ES+DESI BAO), it retains partial stability, i.e. model is not fully stable for the parameter space subjected to  (CC+Pantheon+SH0ES+DESI BAO). In the next section, we discuss in detail the dynamical evolution of the universe as described by the model.

\begin{table}[htbp]
\scriptsize
\centering
\begin{tabular}{|c|c|c|c|c|c|c|c|c|c|c|}
\hline
Dataset & $H_{0}$ [Km/(sec.Mpc)] & $\beta$  & $m$ &  $n$  &  $M$ & $q_{0}$ & $z_{t}$ & $t_{0}$ (Gyr) & $j_{0}$ & $\omega_{eff}$  \\
[1.0ex] \hline
CC & $69.3^{+1.4}_{-1.6}$ & $1.49\pm 0.87$ &$2.55^{+0.43}_{-0.61}$ &  $0.056\pm 0.042$ & - & -0.5570 & 0.637 & $13.51$ & $1.038$ & $-0.704$ \\
[1.5ex] \hline
joint  & $68.58^{+0.95}_{-1.10}$  & $1.51\pm 0.87$  &  $2.18^{+0.28}_{-0.37}$ &  $-0.001\pm 0.03 $ & $-19.386\pm 0.016 $ & $-0.5505$ & $0.727$ & $13.9$  & $0.9614$ & $ -0.700  $ \\
[1.5ex] \hline
\end{tabular}
\caption{ For CC and Joint data, the constrained model parameters with the present values of $q_{0}, z_{t}$, $t_{0}$, $j_0$ and $\omega_0$.}
\label{table:1}
\end{table}
\normalsize
 \section{Physical and dynamical properties of the model}
\label{sec:6} 
\subsection{The physical behaviour of models}
\label{sec:6.2}
We analyse the physical behaviour exhibited by principal quantities like energy density, pressure and EoS parameter. Throughout the entire cosmic evolution, the energy density remains positive, whereas the pressure can become negative in the recent epoch. Such negative pressure emerges precisely during the transition regime from decelerated to accelerated expansion when dark energy becomes the dominant component while the energy density concurrently preserves its positive nature.  
\vspace{0.2cm}\\
Within the constrained parameter values, the model's energy densities increase with redshift $z$ (equivalent to decreasing over cosmic time $t$) and remaining positive values during the entire expansion history. In other words, the energy density will be decreasing from past to future. Significantly, the energy density exhibits its expected positive value, demonstrating its role in driving cosmic expansion. For the median values of the model parameters, the pressure starts out negative in the early universe (at large redshift). It continues to stay negative throughout the present era and into all later epochs. All these features are consistent with the observed accelerating expansion of the universe. These studies are consistent with the accelerating universe’s expanding behaviour.

The EoS parameter provides valuable information about the dominant cosmic fluid and plays a significant role in explaining the universe’s expansion dynamics. In cosmological models, it is often used to determine the nature of dark energy. The EoS parameter characterizes cosmic evolution, including  radiation-dominated phase $(\omega=\frac{1}{3})$, cold dark matter or dust dominated phase $(\omega=0)$, the quintessence dominated regime for $-1 < \omega < -\frac{1}{3}$, the cosmological constant for $\omega = -1$ and phantom-dominated regime $\omega < -1$.
\par
\begin{center}
\begin{figure}
\includegraphics[width=14.5cm, height=9.5cm]{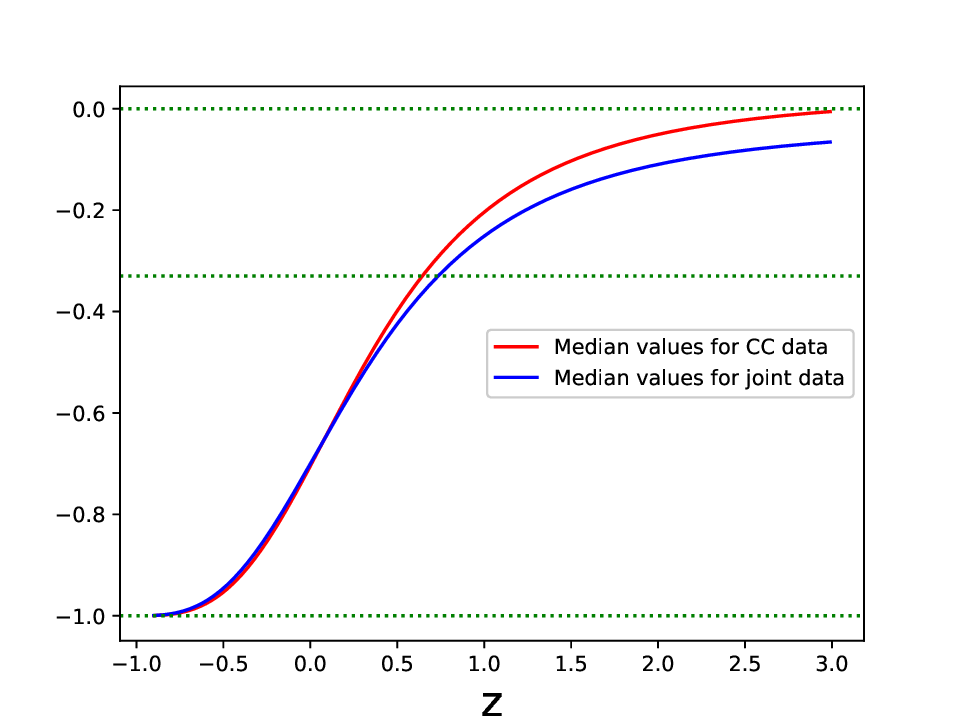}
\caption{$\omega_{eff} $ versus $\mathit{z} $.}
\label{fig:10}
\end{figure}
\end{center}

 Figure $(\ref{fig:10})$ illustrates the evolution of the EoS parameter in the framework of the considered model. At $z=0$,  the EoS parameter for this model is determined to be $\omega_{eff} =-0.704 $ using the CC dataset and $\omega_{eff} = -0.700 $ using the Joint dataset. The evolution of the EoS parameter for this model exhibit quintessence kind of dark energy. This model exhibits cosmological constant-like behaviour in late-time evolution as $z \rightarrow -1$. This model exhibits a matter - dominated expansion phase in the past as evidenced by figure $(\ref{fig:10})$ which shows $\omega = 0$ during early cosmic times.  The cosmic evolution is currently dominated by quintessence like dark energy $-1 < \omega < -\frac{1}{3}$, ultimately approaching the $\Lambda$CDM scenario $\omega=-1$ as $z \rightarrow -1$. The smooth evolution of $\omega_{eff}$ suggests that the model successfully describe the transition from decelerated to accelerated expansion and may provide insights into the dynamical nature of dark energy. The model provides a consistent and observationally viable description of the expansion history of the Universe. At the late time era future  ($z \to -1$) for median values of both data sets $\omega_{eff} \to -1$. Crucially, $\omega_{eff}$  stabilize and do not cross into the phantom domain ($\omega_{eff} < -1$). It implies the universe will eventually approach a stable de Sitter-like expansion phase, effectively avoiding pathological space-time singularities such as the Big Rip. It doesn't violate the Null Energy Condition (NEC) in the future. 
\subsection{Cosmographic parameter}\label{sec:6.4}
\begin{figure}[!htb]
\captionsetup{skip=0.4\baselineskip,size=footnotesize}
   \begin{minipage}{0.40\textwidth}
     \centering
     \includegraphics[width=9.0 cm,height=7.5cm]{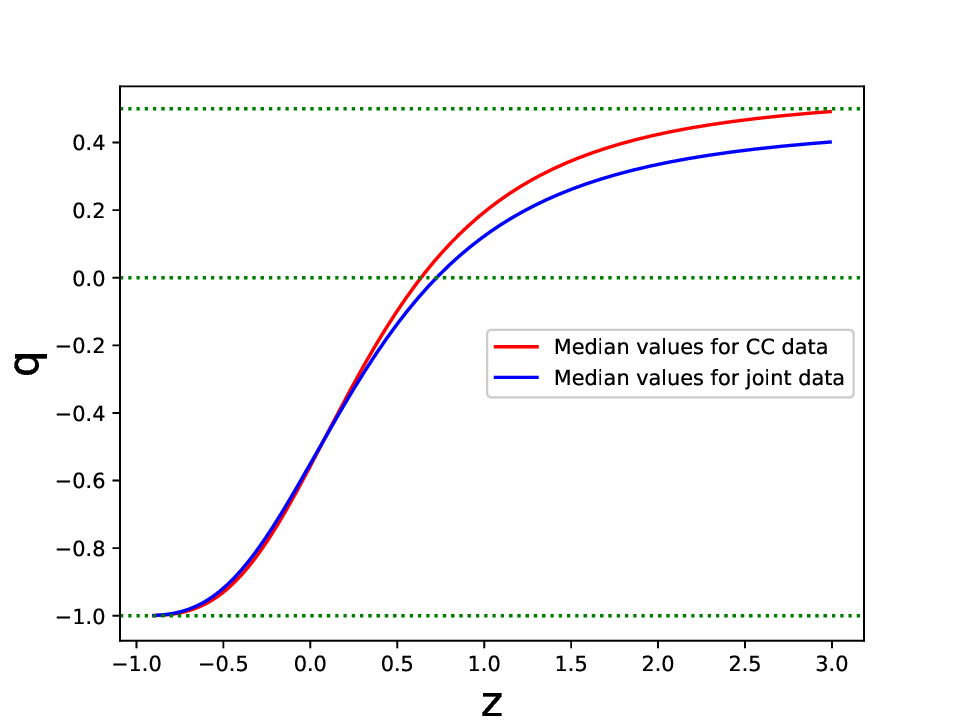}
\caption{ The deceleration parameter versus $z$.}
\label{fig:3}
    \end{minipage}\hfill
   \begin{minipage}{0.40\textwidth}
     \centering
     \includegraphics[width=9.0 cm,height=7.5cm]{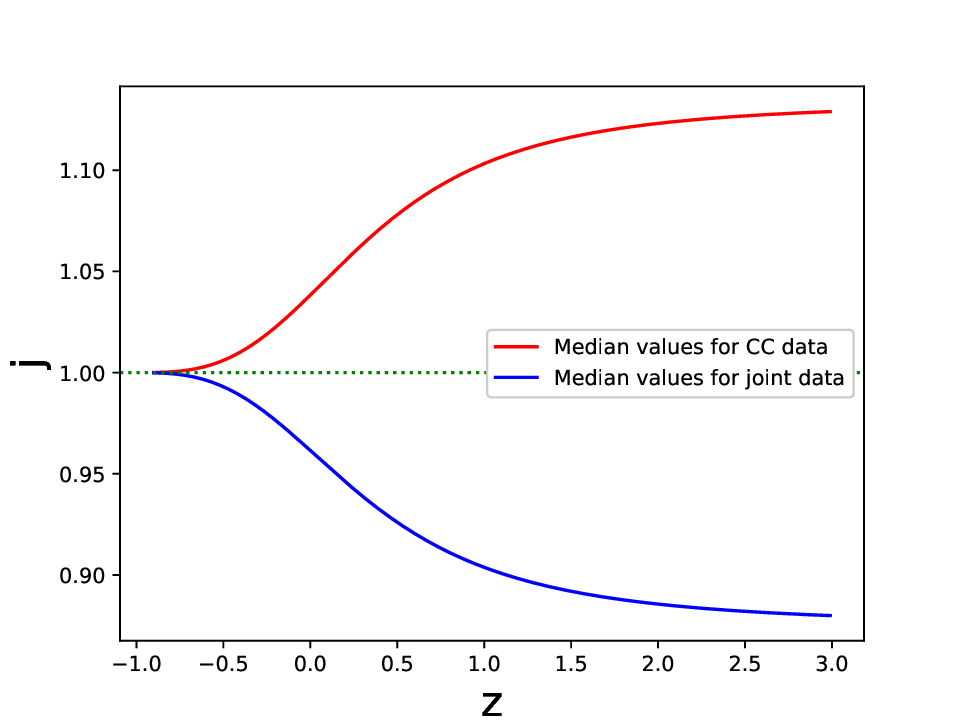}
  \caption{The Jerk parameter versus $z$.}
\label{fig:4}
   \end{minipage}
\end{figure}

This section focuses on analyzing the role of cosmography in cosmology. Weinberg \cite{weinberg2008cosmology} initially introduced the study of cosmography who utilizing a Taylor series to describe the scale factor that emerged around the present time $t_{0}$. Different aspects of cosmographic parameters may be explored in a detailed overview of cosmography  \cite{capozziello2019extended}. In cosmological studies, the Hubble parameter $(H)$ is viewed as a variable observable. The Hubble parameter (\ref{601}) illustrates how the expansion rate of the universe evolves in the given model. The second-order derivative of the scale factor $(a)$ characterizes the evolution of the deceleration parameter $q$ \cite{mukherjee2016parametric}. In the study of the universe’s cosmic evolution, the snap $(s)$ and jerk $(j)$ parameters act as fundamental tools. Among the various cosmological parameters, the deceleration parameter helps in understanding the expansion dynamics of the universe. It is primarily used to measure the rate at which the universe's expansion is slowing down. In mathematical terms, it may be represented through the Hubble parameter and its derivative
\begin{equation}{\label{29}}
 q = -1 - \frac{\dot{H}}{H^2}.
\end{equation}
The expansion dynamics of the universe may be categorized into accelerated or decelerated phases based on this parameter’s values, with $q=0$ indicating the transition phase. When $q > 0$ the universe experiences deceleration, when $q < 0$ it accelerates and a power-law expansion when  $-1 < q < 0$. A super exponential expansion occurs in the universe when $q < -1 $, whereas $q = -1 $ represents the de-Sitter phase \cite{garg2024cosmological,SINGH2024865,doi:10.1142/S0219887822501079,doi:10.1142/S0217751X23501695,singh2024role}.
\vspace{0.3cm}\\
By using (\ref{601}) and (\ref{29}), we may obtain
\begin{equation}
q(z)=-1-\frac{3(n+1)^{2}(8\pi +\beta)(1+z)^{\frac{6(n+1)(8\pi+\beta) }{3\beta-n\beta+16\pi} }     }{m (-4\beta-16\pi)+  \left[    (-3+n^{2}-2n)\beta-16\pi(1+n)  \right](1+z)^{\frac{6(n+1)(8\pi +\beta)   }{3\beta-n\beta+16\pi   } }     }.
\end{equation}
The evolution of the deceleration parameter determined for the constrained of model parameters (see  Table $\ref{table:1}$) is depicted in figure $(\ref{fig:3})$. As indicated in figure $(\ref{fig:3})$, the deceleration parameter $q(z)$ was positive in the early universe but it has transitioned to negative $(q<0)$ at present due to dark energy domination. As shown in figure $(\ref{fig:3})$, the universe shifts from a decelerated expansion phase to an accelerated expansion phase at $z = 0.637$ $(z = 0.727)$ for CC (Joint) data and the existence of accelerated cosmic expansion is observe in the model for $z < 0.637$. The obtained values of the deceleration parameter at present $(z=0)$ are $q_{0}=-0.5570$ and $q_{0}=-0.5505$ for CC and Joint data respectively. These $q_0$ values are very close to $q_0=-0.55$ for the $\Lambda$CDM model \cite{capozziello2019extended}. The deceleration parameter's negative values for this Model indicate that the universe is expanding  with accelerated rate in the present era $(z=0)$. However, the transition redshift remains closely aligned with that of the $\Lambda$CDM model. These results suggest that this Model effectively captures the transition of the Universe from a decelerated expansion phase to the present accelerated regime.
\vspace{0.3cm}\\
As depicted in figure $(\ref{fig:3})$, the evolution of $q(z)$ indicates that this model experience a matter-dominated phase of expansion, confirmed by $q \rightarrow \frac{1}{2}$ in early times. As time progresses, the universe exhibits an accelerated phase $(q<0)$ at present and eventually  approaches the de Sitter phase $(q \rightarrow -1)$ for $z \rightarrow -1$. The observational data confirm that this model accurately describe the accelerated expansion of the universe in the present era. The approach of $q \rightarrow -1$ as $ z \rightarrow -1 $ for median values of model parameters for both data set implies that in the far future, the dark energy will completely dominate the cosmic energy budget and the Universe will enter a phase of de-Sitter like expansion. Importantly, $q(z)$ curve $(\ref{fig:3})$  will not crosses below  $q = -1$, which indicating that this model will not show Super exponential$(q < -1)$ and phantom like behaviour  in future. This is consistent with the findings from the companion EoS plot (figure $\ref{fig:10}$), where $\omega_{eff}(z)$  remains above $-1$  throughout.

\par The rate of change in the universe’s acceleration or deceleration is described by the jerk parameter. The jerk parameter’s sign determines whether the universe’s acceleration increases or decreases. The occurrence of transitional phases in cosmic expansion is signified by a positive jerk parameter. The rate of change of the jerk parameter is determined by the snap parameter. They are defined as follows:  
\begin{equation}{\label{109}}
\mathit{j}=\frac{1}{aH^{3}}\left(\frac{d^{3}a}{dt^{3}}\right),\ \ \mathit{s}=\frac{1}{aH^{4}}\left(\frac{d^{4}a}{dt^{4}}\right).
\end{equation}
In this analysis, equation $(\ref{109})$ can be reformulated in terms of redshift for $j$ and $s$ \cite{wang2009probing}.
\begin{equation}{\label{1099}}
j(z)=\frac{dq}{dz}(1+z)+q(z)\left(1+2q(z)\right), \  \ \  \   s(z)=-\frac{dj}{dz}(1+z)-j(z)\left(2+3q(z)\right).
\end{equation}

Figure $(\ref{fig:4})$ illustrated the jerk parameter’s behavior corresponding to the median values of model parameters. In the figure $(\ref{fig:4})$, At higher redshift values $(z>>1)$ corresponding to the early Universe, the jerk parameter tends to approach a nearly constant value for both data sets. This behavior reflects the dominance of matter in the cosmic energy budget in the early universe and higher order deviations in the expansion rate are suppressed. As the redshift decreases, the jerk parameter shows a gradual variation indicating a transition phase in the expansion history of the Universe. This regime corresponds to the epoch where the Universe evolves from a decelerated expansion phase to the present accelerated phase. The deviation of $J$ from a constant value in this region highlights the influence of dark energy or other dynamical components. In the low redshift limit $(z \approx 0)$, which corresponds to the present epoch, the value of the jerk parameter becomes particularly significant. At the present$(z=0)$, the jerk parameter's value is obtain to be $j_0=1.038$ using the CC dataset whereas the Joint estimates yield $j_0=0.9614$ for this model. Table ($\ref{table:1}$) provides a summary of the present day values of the cosmological parameters. From the trends obtained according to the observational data, a decreasing trend in the jerk parameter's graph is observed with its value approaching $1$ in later time. This behavior indicates that these models deviate from $\Lambda$CDM in the early universe but align with it at later time. The obtained values for the cosmographic parameters are consistent with previously reported results \cite{capozziello2019extended}.

\subsection{The age of Universe}
\label{sec:6.5}
The age of the universe is a crucial factor in evaluating the viability of a model in comparison to the observable universe. The cosmic age $t(z)$ of a cosmological model is computed as a function of $z$ \cite{tong2009cosmic} as
\begin{equation}{\label{34}}
t(z)= \int_{z}^{\infty} \frac{dz}{H(z)(1+z)} \,dz. 
\end{equation}
We compute the given integral numerically to determine the present age of the universe for $z=0$ using the Hubble parameter $H(z)$ (Eq. \ref{601}). For this model, the calculated present age of the universe is $t_{0}$ (at $z=0)=13.51$ Gyr from CC data and $t_{0}=13.9$ Gyr for Joint data, both of which closely match the $ \Lambda$CDM model’s age ($t_{0}=13.79$ Gyr) as determined by Planck results \cite{2020A&A...641A...6P}.
\section{Conclusions}\label{sec:7}
\par In this work, we assess the feasibility of a unified cosmic expansion model within $f(Q, T)$ gravity. In this study, we probe the late-time acceleration of the Universe within the $f(Q, T)$ gravity framework using an affine equation of state  (EoS) $p = n \rho - m$. In our analysis, we considered the functional form $f(Q,T)= Q + \beta T$ with $\beta$ being a free parameter. We utilize CC and Joint (CC+Pantheon+SH0ES+DESI BAO) datasets to perform a Markov Chain Monte Carlo (MCMC) analysis based on the Hubble parameter (\ref{601}). Table (\ref{table:1}) provide a summary of the constrained values obtained from the MCMC analysis for Model (\ref{601}). From figure $(\ref{fig:1})$, it is evident that the model aligns well with the $\Lambda$CDM model for $0 < z < 1$. At large redshift scales $z > 1$, the Hubble parameter behaviour of the model deviates from that of the $\Lambda$CDM model. A comparative visual inspection reveals an excellent statistical concordance between our proposed dynamic Model (represented by the solid black curve) and the observational data across the entire redshift range. Crucially, the theoretical trajectory of our framework maintains a tight geometric overlap with the predictions of the standard $\Lambda$CDM model (dashed red curve). Our model  successfully capture the monotonically increasing expansion rate at higher redshifts.

\par In this model, physical parameter’s behaviour viz (Energy density, Pressure and EoS parameter) have been discussed for median values of model parameters. The energy density remains positive throughout the expansion history while the pressure start with negative values at high redshifts i.e. in the early universe. This negative pressure persists throughout the current epoch and continues into all subsequent cosmic phases due to the dominant of dark energy. These studies are consistent with the observed accelerating expansion of the universe. Based on the constrained values of model parameters, the behaviour of EoS parameter is illustrated in figure (\ref{fig:10}). At present, we obtain the EoS parameter values as $\omega_{eff} = -0.704$ and $\omega_{eff} = -0.700$ for the CC and Joint data respectively in this model. These values at the present time indicate that this model shows a quintessence type of dark energy. Overall, this model with an affine equation of state aligns well with observational data on cosmic acceleration. The evolution of the equation of state parameter $\omega_{eff}$ (figure \ref{fig:10}) demonstrates a clear transition in the dynamical behavior of the Universe.  At high redshift, $\omega_{eff}$ approaches values close to zero, which indicate a matter dominate at early era. As the redshift decreases, $\omega_{eff}$ evolves toward negative values, crossing the threshold required for cosmic acceleration  $(\omega_{eff}< -\frac{1}{3})$ indicating the dominance of dark energy. At the late time epoch, the parameter approaches $\omega_{eff} \approx -1$, consistent with observational constraints and the $\Lambda$CDM model. 

\par Furthermore, we investigate the evolution dynamics in the present study  by examining various cosmological parameters. Moreover, the behavior of the deceleration parameter has been thoroughly discussed. The curve of the deceleration parameter clearly indicates that the universe has transition from a decelerating to an accelerating phase. At present, we obtain the deceleration parameter values as $q_{0}=-0.5570 $ and  $q_{0}=-0.5505 $ based on the CC and Joint data respectively  and the value of transition redshift is $z_{t}=0.637 $ for CC data and $z_{t}=0.727 $ for Joint data. A negative value of the deceleration parameter at $z=0$ indicates that the universe is currently undergoing accelerated expansion. According to deceleration parameter curve (figure $\ref{fig:3}$), the model exhibits a matter-dominated phase in the early era. As time progresses, the universe enters an accelerated expansion phase with $q<0$ at present, eventually approaching the de-Sitter phase where $(q \rightarrow -1)$ for $z \rightarrow -1$. For the median values of model parameters, behaviour of the jerk parameter is illustrated in figure $(\ref{fig:4})$. As the universe evolves, the value of jerk parameter decreases from early to late times and finally approaches to $1$, this demonstrates that the model deviates from the $\Lambda$CDM model in the early universe but aligns with $\Lambda$CDM model  at later times. Additionally, the jerk parameter’s present values are $j = 1.038$ for the CC data and $j = 0.9614$ for the joint data.

\par For the model described by equation (\ref{601}), the current value of Equation of State(EoS) parameter ($\omega$) is close to $-0.7$, while the deceleration parameter values are $-0.557$ for CC data and $-0.5505$ for Joint data. Within this framework, dynamical dark energy is responsible for the observed acceleration of the universe. The model suggests that the cosmic acceleration arises due to a dynamical dark energy. For the median values of the model parameters, the jerk parameter is close to the $\Lambda$CDM limit $(j=1)$. The results on cosmographic parameters are summarized in Table (\ref{table:1}). It is worth mentioning that these quantities will expected to approach the $\Lambda$CDM limits in the far future. Lastly, we estimate the current age of the universe within the framework of this $f(Q, T)$ gravity model. The value of $t$ (at $z=0)$ is $13.51$ Gyr for the CC dataset and $13.9$ Gyr for the joint estimates. Overall, at present times, we find that this  model agrees with current observations of the accelerating universe.

\par For the $\Lambda$CDM model at late times (   $z \rightarrow -1$), the cosmological  parameters $q \rightarrow -1, \omega \rightarrow -1, j \rightarrow 1$ and Snap $(s)\rightarrow 1$. $H$ remains constant and $q=-1$, $j=1$ and $S=1$ in the de-Sitter model of cosmic expansion. This suggests that the universe is experiencing exponential acceleration with the rate of change of both the acceleration and jerk are increasing over time. It is evident that the model closely follows the $\Lambda$CDM behavior and remains consistent with observational data across the entire redshift range. This indicates that the model provides a reliable description of the expansion history of the Universe. The model predictions remain in excellent agreement with both observational data and the $\Lambda$CDM benchmark which indicate that it provides a consistent and viable description of the cosmic expansion history.
  
Overall, it demonstrates that model successfully introduces a time varying, dynamic dark energy component. Consequently, the proposed framework emerges as a highly competitive phenomenological alternative, capable of resolving the dynamical nature of dark energy while fully preserving the empirical triumphs of the standard $\Lambda$CDM cosmological background. The remarkable similarity between the model and $\Lambda$CDM Model reconstructions underscores the statistical robustness of the proposed model in capturing a dynamically evolving universe.  

\section*{\textbf{Acknowledgements}}
GPS is thankful to the Inter-University Centre for Astronomy and Astrophysics (IUCAA), Pune, India for support under Visiting Associateship program. 


\end{document}